\documentclass{aa}
\usepackage{graphics,psfig}
\begin{document}

   \thesaurus{07     % A&A Section 7: Stellar atmospheres
              (02.12.1;  % Line: formation,
               08.01.3;  % Stars: atmospheres,
               08.03.3;  % Stars: chromospheres,
               08.09.2 Sirius;  % Stars: individual,
               08.09.2 Vega  ;  % Stars: individual,
               13.21.5)} % Ultraviolet: stars
   \title{Fe II emission lines in the UV spectrum of Sirius-A and
               Vega\thanks{Based on observations with the NASA/ESA
               {\em Hubble Space Telescope} obtained at the Space
               Telescope Science Institute, which is operated
               by AURA, Inc., under NASA contract NAS 5-26555.}}

   \author{M. van Noort \inst{1,}\thanks{Present address:
               School of Physics, University of Sydney,
	       NSW 2006, Australia }
           \and
           T. Lanz \inst{1}
           \and
           H. J. G. L. M. Lamers \inst{1, 2}
           \and
           R. L. Kurucz \inst{3}
           \and
           R. Ferlet \inst{4}
           \and
           G. H\'ebrard \inst{4}
           \and
           \\ A. Vidal-Madjar \inst{4}
           }
           
   \offprints{T. Lanz}

   \institute{Sterrenkundig Instituut, Utrecht Universiteit,
              Princetonplein 5, NL-3508 TA Utrecht, The Netherlands \\
              (noort@physics.usyd.edu.au, lanz@fys.ruu.nl)
          \and
              SRON Laboratory for Space Research,
              Sorbonnelaan 2, NL-3584 CA Utrecht, The Netherlands
              (hennyl@sron.ruu.nl)
          \and
              Harvard-Smithsonian Center for Astrophysics,
              60 Garden Street, Cambridge MA 02138, USA
              (kurucz@cfa.harvard.edu)
          \and
              Institut d'Astrophysique de Paris, CNRS,
              98 bis Boulevard Arago, F-75014 Paris, France
              (ferlet, hebrard, vidalmadjar@iap.fr)
             }

   \date{Received 29 December 1997; accepted 10 March 1998}
%  \date{Revised 11 February 1998}

   \maketitle

   \begin{abstract}

We present high-quality HST/GHRS spectra in the Hydrogen L$\alpha$
spectral region of Vega and Sirius-A. Thanks to the signal-to-noise
ratio achieved in these observations and to the similarity of the two
spectra, we found clear evidence of emission features
in the low flux region, $\lambda\lambda$1190-1222\,\AA. These emission
lines can be attributed unambiguously to \ion{Fe}{ii} and
\ion{Cr}{ii} transitions.
In this spectral range, silicon lines are observed in absorption.

We built a series of non-LTE model atmospheres with different,
prescribed temperature stratification in the upper atmosphere and
treating \ion{Fe}{ii} with various degrees of sophistication in non-LTE.
Emission lines are produced by the combined effect of the Schuster
mechanism and radiative interlocking, and can be explained
without the presence of a chromosphere. Silicon absorption
lines and the L$\alpha$ profile set constraints on the presence
of a chromosphere, excluding a strong temperature rise in layers
deeper than $\tau_{\rm R} \approx 10^{-4}$.

      \keywords{Stars: atmospheres --
                Stars: chromospheres --
                Stars: Sirius, Vega --
                Line: formation
               }
   \end{abstract}

%________________________________________________________________

\section{Introduction}

On the main sequence, A-type stars are at a juncture point between
hot and cool stars. While hot, massive stars undergo strong mass loss
in fast winds ($\dot{M}\ge 10^{-9}$\,M$_\odot$/yr),
cool stars show chromospheric activity
connected to their subsurface convective layers. Both phenomena apparently
disappear or become much weaker at spectral type A. Many studies have
thus been devoted to the outer layers of A-type stars to search for
indications of a wind or of stellar activity. Several attempts
to detect signatures of weak winds in main-sequence A stars
have been unsuccessful (e.g. Lanz~\& Catala \cite{lanz92}). Recently,
however, a quite weak, blue-shifted absorption was detected in the
\ion{Mg}{ii} resonance lines of Sirius, and interpreted as a wind signature
(Bertin et~al. \cite{sirius3}). A mass loss rate of 
$\dot{M}\approx 10^{-12}$\,M$_\odot$/yr was derived, consistent with the
idea that A-type star winds are radiatively-driven like the winds of
hotter stars. On the cool side, a limit to chromospheric activity
has been set at A7 (B\"ohm-Vitense~\& Dettmann \cite{bohmvitense80},
Marilli et~al. \cite{marilli97}, Simon~\& Landsman \cite{simon97}). 
The most common diagnostics of chromospheres and winds are emission
features. Therefore, we are not expecting emission lines in A stars,
except cases where such lines arise from the circumstellar environment.

High-quality ultraviolet spectra have become available with the
{\em Goddard High Resolution Spectrograph} (GHRS) aboard the
{\em Hubble Space Telescope} (HST). Even the core of strong resonance
lines, including \ion{H}{i} L$\alpha$, can be observed with a reasonably good
signal-to-noise ratio. This makes it possible to investigate
in greater detail the line profile of strong resonance lines. They are
the best tool to probe the outer layers of stars, being
indeed formed very high in the atmosphere.
In this respect, L$\alpha$ is most interesting because it spans the largest
range of depth of formation, from the far wing to the line core.
This large variation in opacity also affects the formation of lines of other
elements, especially close to L$\alpha$ core. Such lines see 
a much lower local pseudo-continuum than lines outside L$\alpha$,
and will be formed much higher in the atmosphere than weak lines in
other regions.

%_____________________________________________________

\begin{table*}
  \caption[]{Observation log.}
  \label{TabObs}
  \begin{tabular}{lccccr}
  \hline
  &&&&& \\ [-3mm]
  Target & Spectral Range & Date of Observation & GHRS Grating & Aperture &
  Exposure time \\
  \hline
  &&&&& \\ [-3mm]
Sirius-A & 1188 \AA\ -- 1218 \AA & 1996 Nov 20 & G140M & SSA & 1632.0 s \\
Sirius-A & 1278 \AA\ -- 1307 \AA & 1996 Nov 20 & G140M & SSA & 217.6 s \\
Sirius-A & 1308 \AA\ -- 1337 \AA & 1996 Nov 20 & G140M & SSA & 217.6 s \\
Vega & 1185 \AA\ -- 1222 \AA & 1996 Dec 23 & G160M & SSA & 435.2 s \\
Vega & 1274 \AA\ -- 1311 \AA & 1996 Dec 23 & G160M & SSA & 108.8 s \\
Vega & 1303 \AA\ -- 1341 \AA & 1996 Dec 23 & G160M & SSA & 108.8 s \\
  \hline
  \end{tabular}
\end{table*}
%_____________________________________________________

In Sect. 2 and 3, we will describe our GHRS observations around L$\alpha$
of two bright A stars, Vega and Sirius-A. We will in particular point
out the presence of \ion{Fe}{ii} and \ion{Cr}{ii} emission lines between
1190 and 1222\,\AA.
Bertin et~al. (\cite{sirius2}) noticed the presence of emission features
around L$\alpha$ in a Cycle~1 GHRS spectrum of Sirius-A, originally recorded
to derive the D/H abundance ratio in the local interstellar medium. This
prompted us to repeat and extend these observations to investigate
their origin. An explanation of these emission features is given in the
second half of the paper. In Sect.~4, we describe our new non-LTE model
atmospheres. We explore and set limits on a chromosphere (Sect.~5),
and investigate non-LTE effects in \ion{Fe}{ii} line formation (Sect.~6).

%_____________________________________________________

\section{Observations and data reduction}

\subsection{HST observations}

The far ultraviolet observations of the very bright stars Sirius and Vega were 
made in the framework of our Cycle~6 Guest Observer proposals ID6800 and
ID6828 with the Goddard High Resolution Spectrograph aboard HST.
A first attempt was performed in September 1996 but failed because the stars 
were not correctly located within the GHRS entrance slit. The observations
were repeated in November and December 1996 with a different pointing
strategy which was then fully successful. We have obtained
high-quality spectra of both stars at
high and medium spectral resolution. Only medium resolution spectra
are discussed in this paper.
Table~\ref{TabObs} lists the individual observations.

%______________________________________________ 
\begin{figure*}
  \resizebox{18cm}{!}{\includegraphics{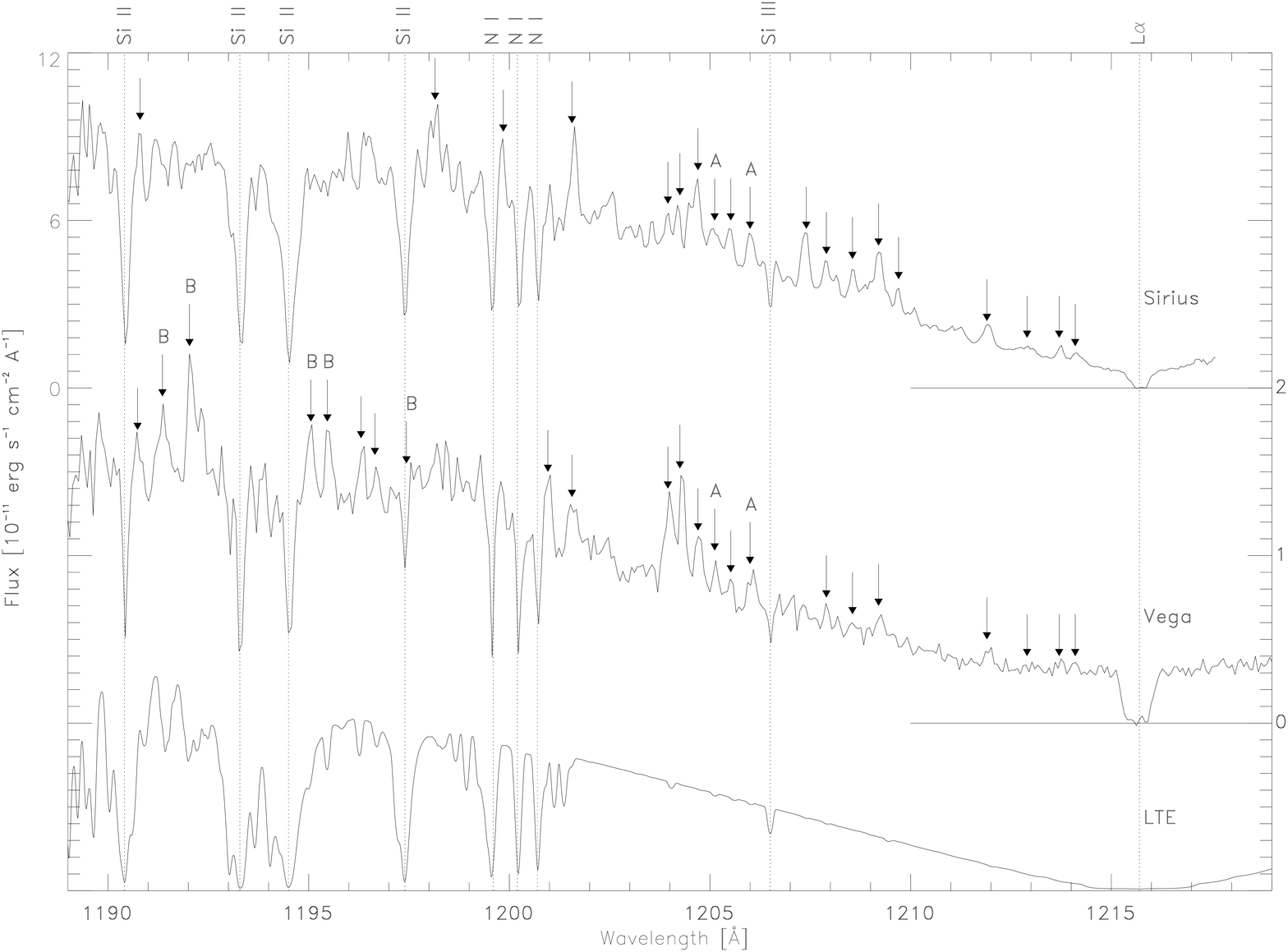}}
%  \rule{0.4pt}{2cm}
  \caption{GHRS spectra of Sirius-A (top) and Vega (middle), together with a
     LTE spectrum (bottom). The flux scale is given on the left for Sirius,
     and on the right for Vega. Zero-flux level is also indicated by
     horizontal lines. For better clarity, the LTE spectrum has been
     shifted down. Identifications of the strongest absorption lines
     are shown with dotted lines, while the strongest emission lines
     are indicated by arrows. For detailed identifications, see
     Table \ref{TabIdent}. A few lines which will be modeled in detail
     (Sect.~6) are labelled as A and B.}
  \label{svegspc}
\end{figure*}
%_____________________________________________________

In the first order, the G140M and G160M gratings provide a
resolving power of $\simeq$\,20\,000 and $\simeq$\,16\,000 respectively at
1200\,\AA, {\it i.e.} a
spectral resolution of about 15\,km/s and 19\,km/s in this spectral range. 
The resolution is slightly better in the other spectral ranges. In order to
keep the best possible resolution, we have used only the Small Science Aperture 
(SSA) which corresponds to 0.25" on the sky and illuminates one diode. Further 
details on the instrumentation can be found in Heap et~al. (\cite{ghrs}).

We used the FP-SPLIT mode which splits the total exposure time into successive 
cycles of four sub-exposures, each corresponding to a slightly different 
projection of the spectrum on the photocathode. We also used the ``quarter 
stepping'' mode which provides a sample of four pixels per resolution element. 
This allows to simultaneously oversample the spectrum (the 
SSA does not fulfill the Nyquist sampling criterion) and to correct for the 
granularity of the photocathode. The granularity is identical
for the four sub-exposures, and can therefore be estimated from
a comparison of these sub-exposures which have different
photon statistical noise.

\subsection{Data reduction}

The data were reduced with the IRAF software, using the STSDAS package.
We used the standard method for correcting 
for the granularity, which is available in the IDL package 
({\it corre\_hrs} procedure). We found that this procedure allows us to 
achieve signal-to-noise ratios consistent with the high data count rates.
Without rebinning our spectra ({\it i.e.} they show four pixels per resolution
element), the S/N ratio of the final summed spectrum is larger than
100:1 in the long wavelength range and of the order of 50:1 and 15:1
in the L$\alpha$~range, for Sirius and Vega respectively. At the bottom
of L$\alpha$ near the narrow interstellar absorption core (see Fig.~1),
the S/N ratio is about 14:1 and 7:1 respectively. The geocoronal
L$\alpha$ emission has been removed, as described in Bertin et~al.
(\cite{sirius2}).

In radial velocity, the absolute calibration accuracy is the standard one,
as provided by the GHRS pipeline, {\it i.e.} $\pm$3.0 km/s. After the
standard reduction, we shifted Vega and Sirius spectra by 
0.05 and 0.02 \AA\  respectively, based on the 
identification of the \ion{N}{i}, \ion{Si}{ii} and \ion{Si}{iii} absorption 
lines. This shift is in good agreement with the radial velocities listed in
the Simbad database, operated at CDS.
%_____________________________________________________

\section{Qualitative description of the spectra}

In this section, we compare qualitatively the spectrum of \object{Vega}
and \object{Sirius}. We compare them
also to LTE theoretical spectra which
have been computed with the synthetic spectrum
program SYNSPEC (Hubeny et al. \cite{synspec}), using
Kurucz (\cite{kurucz91}) LTE line-blan\-keted model atmospheres. We have
adopted the following stellar parameters: $T_{\rm eff}\,=\,9400$\,K,
$\log g\,=\,3.95$, and a metallicity one third solar for Vega;
$T_{\rm eff}\,=\,9900$\,K, $\log g\,=\,4.05$, and a metallicity three
times solar for Sirius-A. These parameters differ slightly from values
adopted by other authors (Castelli~\& Kurucz \cite{castelli94},
Hill~\& Landstreet \cite{ghill93}) which are primarily based on the study
of the visible spectrum, but they provide a good fit to the
observations in the $\lambda\lambda$1270--1340\,\AA\  region. We do not
expect however that these LTE spectra match the observations
in the L$\alpha$ region well (see Hubeny \cite{hubeny81}), 
but they should prove useful for the purpose of line identification. We will
now describe in turn the two observed spectral regions.

\subsection{The L$\alpha$ region}

\begin{figure*}
  \resizebox{18cm}{!}{\includegraphics{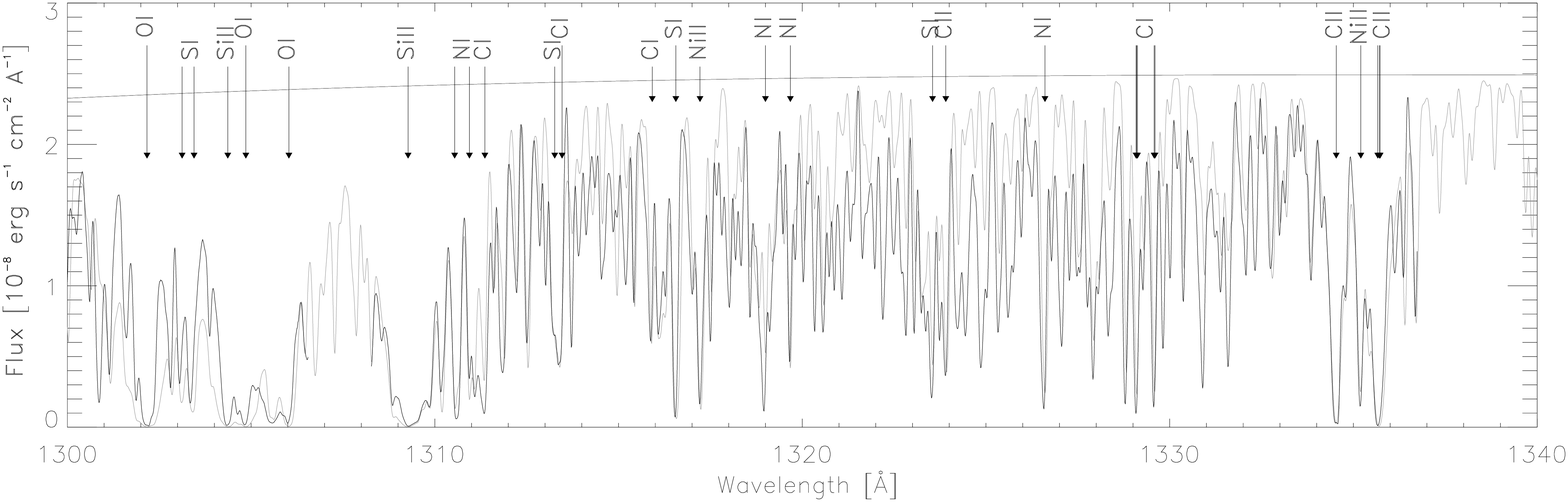}}
  \resizebox{18cm}{!}{\includegraphics{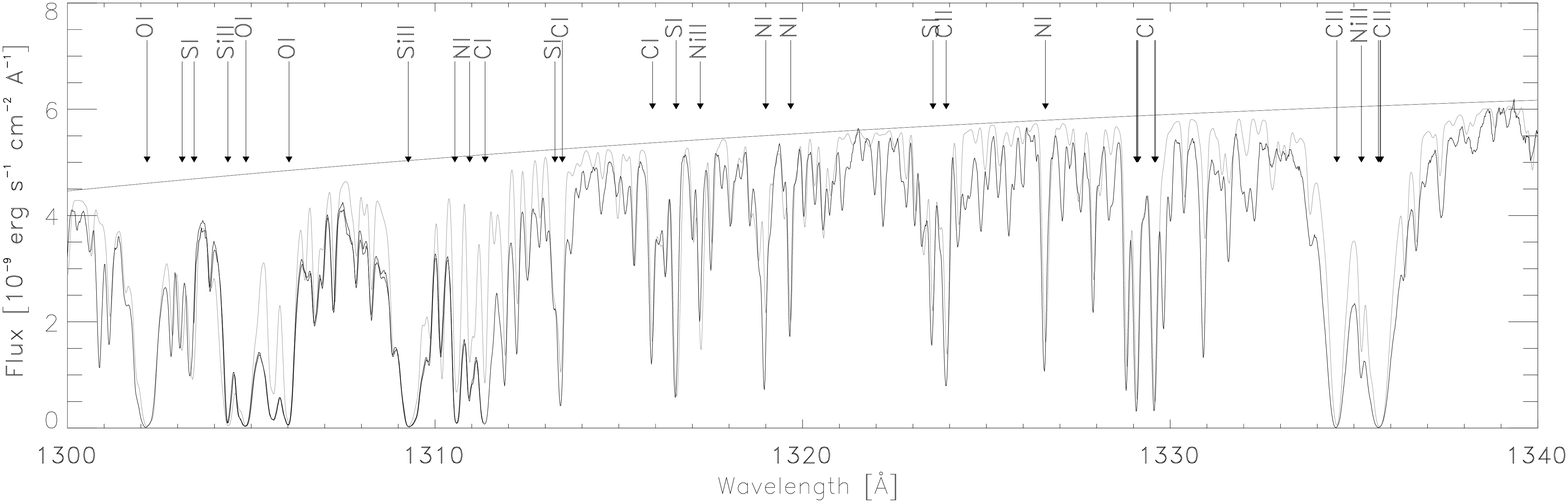}}
%  \rule{0.4pt}{2cm}
  \caption{GHRS spectra of Sirius (top) and Vega (bottom), and 
          LTE theoretical spectra and continua (dotted lines).}
  \label{spec13}
\end{figure*}

The blue wing and the core of L$\alpha$ are displayed in Fig.~\ref{svegspc},
after rebinning to the nominal spectral resolution.
In this figure, the LTE spectrum corresponds to the best fitting model
of Vega in the $\lambda$1300\,\AA\  spectral region (see Fig.~\ref{spec13}).
As expected, this LTE model does not fit well the L$\alpha$ blue wing if one
assumes the scaling factor determined at 1320\,\AA. Since we use it here
only for illustration and line identification purposes, we have
multiplied the theoretical flux by an additional factor 2 to have a scale
similar to the observed spectra in the L$\alpha$ region.

%
%___________________________________ Two column table (place early!)
   \begin{table*}
      \caption[]{Identified lines in Vega and Sirius-A spectra,
                  including an indication if the line is in
                  absorption (abs.) or in emission (em.).
                  Terms are listed only for \ion{Fe}{ii} and
                  \ion{Cr}{ii} transitions.
                  A few emission lines remain unidentified, but are
                  listed for completeness.}
      \label{TabIdent}
      \begin{tabular}{c c c l r r r r r}
      \hline
       &&&&&&&& \\ [-3mm]
       Wavelength&Vega&Sirius-A&Ion&Lower energy&Term&Upper energy&Term&
            $\log gf$ \\
       \multicolumn{1}{c}{[\AA]} &&&& [cm$^{-1}$] && [cm$^{-1}$] && \\
      \hline
       &&&&&&&& \\ [-3mm]
       1190.412 & abs. & abs. & \ion{Si}{ii} & 0.000 & & 84004.523 & & 0.120 \\
       1190.736 & em. & em. &&&&&& \\
       1191.356 & em. & -- & \ion{Fe}{ii} & 8391.938 & $a~^4$D & 
                                 92329.891 & $v~^4$F$^{\rm o}$&-1.868 \\
       1192.030&em.& -- &\ion{Fe}{ii}&8391.938&$a~^4$D&
                                 92282.461&$v~^4$F$^{\rm o}$&-1.448\\
       1193.009&abs.&abs.&\ion{C}{i}&16.400&&83838.078&&-0.740\\
       1193.286&abs.&abs.&\ion{Si}{ii}&0.000&&83802.211&&0.420\\
       1194.445&em.& -- &\ion{Fe}{ii}&25428.783&$a~^4$G&
                                     109149.680&$x~^2$I$^{\rm o}$&-1.181\\
       1194.498&abs.&abs.&\ion{Si}{ii}&287.320&&84004.523&&0.820\\
       1195.055&em.& -- &\ion{Fe}{ii}&8680.454&$a~^4$D&
                                     92358.609&$v~^4$F$^{\rm o}$&-1.909\\
       1195.465&em.& -- &\ion{Fe}{ii}&8680.454&$a~^4$D&
                                     92329.891&$v~^4$F$^{\rm o}$&-1.509\\
       1196.263&em.& -- &\ion{Fe}{ii}&8680.454&$a~^4$D&
                                     92274.117&$u~^4$P$^{\rm o}$&-1.646\\
       1196.670&em.& -- &\ion{Fe}{ii}&21251.607&$a~^4$H&
                                     104816.797&$w~^4$H$^{\rm o}$&-1.088\\
       1197.391&abs.&abs.&\ion{Si}{ii}&287.320&&83802.211&&0.120\\
       1197.435&em.& -- &\ion{Fe}{ii}&8846.768&$a~^4$D&
                                    92358.609&$v~^4$F$^{\rm o}$&-1.585\\
       1197.498&em.& -- &\ion{Fe}{ii}&21430.359&$a~^4$H&
                                    104937.797&$w~^4$H$^{\rm o}$&-1.020\\
       1198.087&em.&em.&\ion{Fe}{ii}&25805.328&$a~^4$G&
                                   109271.711&$x~^2$I$^{\rm o}$&-1.751\\
       1198.366&em.&em.&\ion{Fe}{ii}&21581.639&$a~^4$H&
                                   105028.602&$w~^4$H$^{\rm o}$&-1.201\\
       1198.931&em.&em.&\ion{Fe}{ii}&21251.607&$a~^4$H&
                                   104659.258&$w~^4$H$^{\rm o}$&0.162\\
       1199.236&em.&em.&\ion{Fe}{ii}&21430.359&$a~^4$H&
                                   104816.797&$w~^4$H$^{\rm o}$&0.017\\
       1199.550&abs.&abs.&\ion{N}{i}&0.000& &83364.617& &-0.290\\
       1199.671& -- &em.&\ion{Fe}{ii}&21581.639&$a~^4$H&
                                    104937.797&$w~^4$H$^{\rm o}$&-0.112\\
       1200.223&abs.&abs.&\ion{N}{i}&0.000& &83317.828& &-0.460\\
       1200.710&abs.&abs.&\ion{N}{i}&0.000& &83284.070& &-0.760\\
       1200.893&em.&em.&\ion{Fe}{ii}& 3117.461&$a~^4$F&
                                     86388.820&$v~^4$D$^{\rm o}$&-1.828\\
       1201.415&em.&em.&\ion{Fe}{ii}&21581.639&$a~^4$H&
                                    104816.797&$w~^4$H$^{\rm o}$&-1.148\\
       1201.506&em.&em.&\ion{Fe}{ii}&21430.359&$a~^4$H&
                                    104659.258&$w~^4$H$^{\rm o}$&-1.295\\
       1201.549&em.&em.&\ion{Fe}{ii}&21711.918&$a~^4$H&
                                    104937.797&$w~^4$H$^{\rm o}$&-1.286\\
       1204.020&em.&em.&\ion{Fe}{ii}&25428.783&$a~^4$G&
                                    108483.867&$t~^4$G$^{\rm o}$&-1.279\\
       1204.250&em.&em.&&&&&& \\
       1204.755& em.&em.&\ion{Fe}{ii}&25805.328&$a~^4$G&
                                     108809.789&$v~^4$H$^{\rm o}$&-1.601\\
       1205.117&em.&em.&\ion{Fe}{ii}&26170.182&$b~^2$H&
                                    109149.680&$x~^2$I$^{\rm o}$&-0.755\\
       1205.558&em.&em.&\ion{Fe}{ii}&26055.424&$a~^4$G&
                                    109003.297&$^4$D$^{\rm o}$&-1.776 \\
       1205.997&em.&em.&\ion{Fe}{ii}&26352.766&$b~^2$H&
                                    109271.711&$x~^2$I$^{\rm o}$&-0.780\\
       1206.500&abs.&abs.&\ion{Si}{iii}&0.00& &82884.406& &0.230\\ 
       1207.360&em.&em.&\ion{Fe}{ii}& 7955.299&$a~^4$D&
                                     90780.617&$v~^4$F$^{\rm o}$ &-2.741\\
       1207.900&em.&em.&\ion{Fe}{ii}&25428.783&$a~^4$G&
                                    108217.570&$p~^4$F$^{\rm o}$ &-0.697\\
       1208.559&em.&em.&\ion{Fe}{ii}&27314.922&$a~^2$F&
                                    110064.953&$q~^2$G$^{\rm o}$ &-2.308\\
       1209.200&em.&em.&&&&&& \\
       1209.700& -- &em.&&&&&& \\
       1211.986&em.&em.&\ion{Fe}{ii}&8391.938&$a~^4$D&
                                    90901.125&$v~^4$P$^{\rm o}$&-1.618\\
       1212.966&em.&em.&\ion{Fe}{ii}&7955.299&$a~^4$D&
                                    90397.867&$u~^4$D$^{\rm o}$&-1.466\\
       1213.759&em.&em.&\ion{Fe}{ii}&8391.938&$a~^4$D&
                                    90780.617&$v~^4$F$^{\rm o}$&-1.174\\
       1214.150&em.&em.&\ion{Fe}{ii}&8846.768&$a~^4$D&
                                    91208.891&$v~^4$F$^{\rm o}$&-1.494\\
       1217.140&--&em. &\ion{Cr}{ii}&12303.860&$a~^6$D&
                                     94656.242&$x~^6$D$^{\rm o}$&-1.138\\
       1218.626&--&em. &\ion{Cr}{ii}&12303.860&$a~^6$D&
                                     94363.508&$  ^6$P$^{\rm o}$&-0.887\\
       1218.906&--&em. &\ion{Cr}{ii}&11961.810&$a~^6$D&
                                     94002.563&$  ^6$P$^{\rm o}$&-1.085\\
       1219.563&--&em. &\ion{Cr}{ii}&12147.820&$a~^6$D&
                                     94144.430&$ ^6$P$^{\rm o}$&-0.939\\
       1219.959&--&em. &\ion{Cr}{ii}&12032.580&$a~^6$D&
                                     94002.563&$ ^6$P$^{\rm o}$&-1.057\\
       1220.076&--&em. &\ion{Cr}{ii}&12303.860&$a~^6$D&
                                     94265.992&$x~^6$D$^{\rm o}$&-1.076\\
       1220.165&--&em. &\ion{Cr}{ii}&12496.440&$a~^6$D&
                                     94452.570&$x~^6$D$^{\rm o}$&-1.067\\
       1220.252&--&em. &\ion{Cr}{ii}&12147.820&$a~^6$D&
                                     94098.133&$x~^6$D$^{\rm o}$&-1.396\\
       1221.492&--&em. &\ion{Cr}{ii}&12496.440&$a~^6$D&
                                     94363.508&$ ^6$P$^{\rm o}$&-0.939\\
      \hline
      \end{tabular}
   \end{table*}
%_____________________________________________________

In both stars, the narrow absorption which reaches zero-flux 
in L$\alpha$ core is due to interstellar absorption, hidding the
stellar Doppler core. Contrary to
the LTE model prediction, the stellar line profiles do not reach
zero-flux. Additionally, the profile differs between the
two stars. The Sirius profile runs deeper towards the line center, 
while the Vega profile flattens out (except for the interstellar
core) between 1212 and 1220\,\AA\  at a flux of
$3\cdot 10^{-12}$\,erg/s/cm$^2$/\AA. This may be a first hint
that a chromosphere is present in Vega, or this may simply result from
different non-LTE heating in the Lyman continuum (see also Sect.~\ref{chrom}).

The second striking feature in these spectra is the presence of numerous
emission lines. Since emission lines have not been reported before in the
spectrum of either star, we shall first examine the reality of these emission
features, then identify them, and finally discuss their origin. While
the presence of emission lines might seem unexpected in main-sequence,
early A-type stars, Hubeny (\cite{hubeny82}) pointed out that some lines
may indeed turn into emission in the L$\alpha$ wings under some
appropriate conditions.

%______________________________________________ 
%______________________________________________ 

The signal-to-noise ratio achieved in these GHRS observations
(see Sect.~2) indicates that we cannot attribute these features to noise.
Such a statement would not have been possible from {\sl IUE\/} spectra.
The similarity of the two spectra, as well as the presence
of these emissions in an earlier GHRS spectrum of Sirius recorded in
August 1992 (Bertin et~al. \cite{sirius2}),
provides additional support to the reality of these emission features. 

The question is therefore to determine if these features are real emission
lines, or if they are simply high points between blends of absorption lines.
Some strong absorption lines can be readily identified: \ion{N}{i}
(1199.5, 1200.2, and 1200.7\,\AA), \ion{Si}{ii} (1194.5, 1193.3,
1197.4, and 1190.4\AA), and \ion{Si}{iii} (1206.5\AA).
Except for the \ion{Si}{iii} line, there are
however no strong absorption lines in the range between 1202 and 1215\,\AA\ 
as verified by the LTE spectrum synthesis. Moreover, there is a very
good correlation between the observed emission features in Vega and Sirius,
and predicted \ion{Fe}{ii} absorption lines in the LTE spectrum.
Therefore, we believe that the explanation that these high points are
only gaps between absorption blends is quite unlikely.

Our confidence in attributing the emission features to \ion{Fe}{ii} emission
lines is greatly increased by the fact that we can identify 
lines from the same multiplet in several instances. A first interesting
case is the doublet $b~^2$H -- $x~^2$I$^{\rm o}$ at 1205.1, 1206.0\,\AA\
(labelled A in Fig.~\ref{svegspc}).
The two lines appear in emission in the spectrum of the two stars.
A second case (B) is the quartet $a~^4$D -- $v^4$F$^{\rm o}$. Four lines
($\lambda$1191.4, 1192.0, 1195.1, 1195.5\,\AA) are in emission in the
spectrum of Vega. A fifth line ($\lambda$1197.4\,\AA) coincides
with the strong \ion{Si}{ii} $\lambda$1197.7 absorption line.
We notice however that this latter line appears weaker than expected
compared to the two other \ion{Si}{ii} lines of the same multiplet. We attribute
this weakness to some filling by the \ion{Fe}{ii} emission line.
Most interestingly, none of these five lines appear in emission in
Sirius. This provides an indirect confirmation of the identification,
and also indicates differences in \ion{Fe}{ii} line formation between the
two stars. We will later use these two multiplets as test cases to study the
origin of the emission lines.

In the red wing of L$\alpha$ (up to 1222\,\AA), there are no obvious
emission lines in the spectrum of Vega, but there are two strong emission
lines at 1220.1 and 1221.5\,\AA\  in the original spectrum of Sirius
recorded in 1992. We can unambiguously attribute these lines to
two \ion{Cr}{ii} multiplets.

Table \ref{TabIdent} lists the identified lines in this region, as well
as four unidentified emission features. With a good certainty,
most emission lines can be attributed to \ion{Fe}{ii} or to \ion{Cr}{ii}.
We do not find any \ion{Fe}{ii} absorption lines in the range, 
$\lambda\lambda$1190--1222\,\AA. Conversely, no emission lines are
found outside this range (see below).

Finally, we have measured in both stars a small wavelength shift
of the order of 5\,km\,s$^{-1}$ of the resonance absorption lines.
We attribute this shift to the deformation
of the line profile due to an interstellar contribution. On the other hand,
we do not find any systematic shift of the emission lines which may have
suggested that these lines were formed in a wind.

\subsection{The $\lambda$1300\,\AA\ region}

Fig.\ \ref{spec13} presents the GHRS spectra in a spectral region near
the flux maximum. The stellar parameters have been set to provide the
best fit to the observations in this range with a LTE spectrum.
In this range, the LTE approximation is reasonably good (except perhaps
for the \ion{Si}{ii} and \ion{C}{ii} resonance lines).

Since Sirius' photosphere has a higher heavy elem\-ent content
(three times solar), its spectrum
is much more depressed relative to the theoretical continuum compared to
Vega (one third solar metallicity). High points in the Vega spectrum
are often close to the theoretical continuum, but always below the continuum.
There is thus no evidence of emission features in this range, where
the spectrum can be well understood with absorption lines only.

Finally, we notice that the \ion{C}{ii}
$\lambda\lambda$1334.5, 1335.7 resonance lines are weaker in Sirius than in
Vega. \ion{C}{i} lines in this range are also weaker, though less noticeably.
To fit the carbon lines in Sirius, we have adopted an abundance,
[C/H]=-0.6\,dex, that agrees well with the carbon deficiency
noted by previous authors (e.g. Lambert et~al. \cite{lambert82}).

\subsection{Summary of observation results}

Thanks to the high quality of GHRS spectra of Vega and Sirius,
we found evidence of
emission features in the blue wing of L$\alpha$. In order to understand
their origin, let us first summarize the properties of these features:
\begin{enumerate}
\item[{\sl i)\/}] The emission features are found in the range,
  $\lambda\lambda$1190--1222\,\AA . No emission lines have been
  detected outside this range (see also e.g. Freire-Ferrero et al.
  \cite{freire83});
\item[{\sl ii)\/}] The emission features can be assigned to
  \ion{Fe}{ii} and to \ion{Cr}{ii} transitions;
\item[{\sl iii)\/}] There is a remarkable resemblance between Vega
  and Sirius spectra, most emission features being found in both spectra.
  Generally, \ion{Fe}{ii} emission lines are somewhat stronger in Vega;
\item[{\sl iv)\/}] The spectrum of Sirius is almost identical to the
  GHRS spectrum obtained by Bertin et~al. (\cite{sirius2}) in 1992.
\end{enumerate}
Finally, a model of Vega and Sirius outer layers must also explain
the significant flux levels observed in the central region of L$\alpha$.

Based on these facts, we can exclude already some possible causes.
While Sirius has a white dwarf companion, we can most likely rule out
any effect from the companion, because the spectra are so similar and
Vega is a single star. Moreover, any contribution from Sirius-B should
remain negligible (Bertin et~al. \cite{sirius2}).
Time dependent phenomena, like transient mass-loss episodes, are also
unlikely due to the absence of variations. Emission lines from a circumstellar
shell would be also present at other wavelengths. Emission features can arise
from a chromosphere or can result directly from non-LTE effects in line
formation. We will investigate both possibilities with new model atmospheres
for early A-type stars incorporating appropriate physics inputs. Since
both spectra are so similar, we will concentrate on modeling the Vega spectrum
and we shall discuss briefly in the concluding section the differences between
the two stars.

%______________________________________________________________

\section{Model atmospheres}

\subsection{General approach}

L$\alpha$ is formed in an extended part of the stellar atmosphere because
of the large increase in opacity from the continuum to the line core.
To compute L$\alpha$ profiles correctly, we need therefore model
atmospheres extending up to very-high, low-density layers. Usually,
Kurucz LTE model atmospheres (Kurucz \cite{kurucz91})
stop before getting to these layers. In any case,
departures from LTE are quite likely in these layers due to the very
low densities, and thus there is a preponderance of radiative processes over
collisional processes. In deeper layers where the continuum is formed, 
the crucial effect is backwarming due to line-blanketing, and non-LTE
effects play a minor r\^ole. Ideally, we would like thus to model Vega and
Sirius with consistent, non-LTE, line-blanketed atmospheres.
We decided however to follow a more empirical approach by choosing
various temperature structures in the outer layers,
and thus to explore what atmospheric structures are compatible with the
GHRS observations.

Our model atmospheres are set up in two pieces. In the deep part, we adopt
a LTE line-blanketed model atmosphere, while we prescribe the temperature
structure in the higher layers. Either we extrapolate the Kurucz temperature
structure, or we assume a temperature plateau or a rise. We assume
hydrostatic equilibrium everywhere to derive the total pressure. In a
second step, we keep the temperature and pressure stratification fixed,
and solve the statistical equilibrium for hydrogen and a few neutral
elements which are important contributors to the continuum opacity
(mainly, \ion{C}{i} and \ion{Si}{i}). To model L$\alpha$, we have
compared the emergent spectrum assuming either complete photon
redistribution in frequency (CRD) or coherent scattering (or partial
redistribution, PRD). Finally, we solve the statistical
equilibrium for \ion{Fe}{ii} to study the non-LTE line formation problem.
The next sections describe in more detail these steps.

\subsection{LTE line-blanketed photospheres}

We computed a series of LTE, line-blanketed model atmospheres
with the ATLAS9 code (Kurucz \cite{kurucz91}). For each model,
we have calculated a LTE spectrum synthesis in the range,
$\lambda\lambda$1270-1340\,\AA. The best match to Vega GHRS spectra
was obtained for the stellar parameters, $T_{\rm eff}\,=\,9400$\,K,
$\log g\,=\,3.95$, and one third solar metallicity. In all models,
we have assumed a turbulent velocity, $V_t=2$\,km\,s$^{-1}$.
All extended models will be built upon this model photosphere.

\subsection{Non-LTE extended model atmospheres}

%______________________________________________ 
\begin{figure}
  \resizebox{8.8cm}{!}{\includegraphics{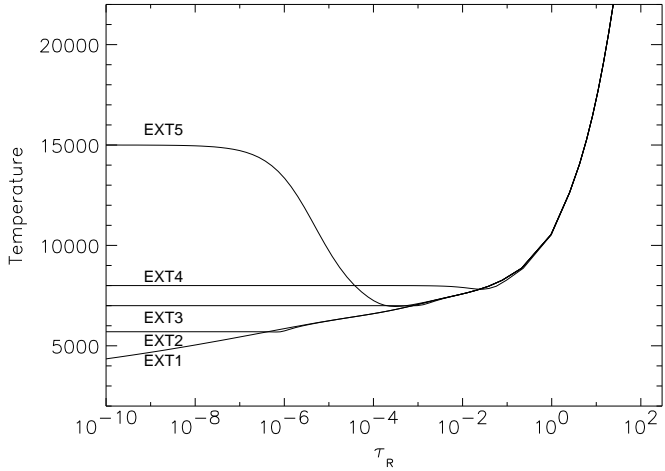}}
%  \rule{0.4pt}{2cm}
  \caption{Temperature stratification of extended model atmospheres.}
  \label{figatmos}
\end{figure}
%_____________________________________________________

We had then to extend the ATLAS9 model atmosphere to smaller optical
depths. We note in passing that the geometrical extension of the atmosphere
itself remains small; a plane-parallel geometry is still adequate.
We extrapolated the temperature stratification from the last depth points
at $\tau_{\rm R}\approx\,10^{-6}$
to depths $\tau_{\rm R}\approx\,10^{-10}$. We considered several cases
(see Fig.~\ref{figatmos}):
\begin{itemize}
\item[{\sl i)\/}] Cool external layers, by extrapolating logarithmically
the T vs. $\tau_{\rm R}$ relation from the ATLAS9 model
(model EXT1);
\item[{\sl ii)\/}] Isothermal outer layers, with various surface temperatures;
note that the isothermal region extends deeper if we choose higher surface
temperature, and that this will eventually change drastically the
overall flux distribution (models EXT2 to EXT4, with boundary
temperatures, $T_0$\,=\,5700, 7000, 8000\,K, respectively);
\item[{\sl iii)\/}] Chromospheric temperature rise (model EXT5).
\end{itemize}
The model with a temperature rise is characterized by three parameters,
the surface temperature, the depth where the temperature starts to rise,
and the zone in which the temperature increases. Many combinations are
possible. The five models that we have selected here will however allow
us to discuss the essential behavior  of L$\alpha$ and \ion{Fe}{ii} lines.
Assuming hydrostatic equilibrium, the total pressure and densities are
then extrapolated straightforwardly.

For each model, EXT1 to EXT5, we derive non-LTE populations for hydrogen
with the TLUSTY194 program (Hubeny \cite{tlusty88},
Hubeny~\& Lanz \cite{clali1}). TLUSTY is a quite versatile program, in
particular allowing one to solve only statistical equilibrium (and radiative
transfer) while keeping the atmospheric stratification (temperature, total
density) fixed.
Hydrogen is represented essentially exactly with the lowest 8 levels included
individually while the upper levels are merged into a non-LTE superlevel
accounting for level dissolution (Hubeny et al. \cite{hhl94}). We have
also included in the non-LTE model a few levels of heavier species which
contribute to the continuum opacity in the UV (\ion{C}{i}, \ion{Mg}{i},
\ion{Al}{i}, \ion{Si}{i}, and \ion{Fe}{i}).

\subsection{CRD and PRD approximations}

%_____________________________________________________
\begin{figure}
  \resizebox{8.8cm}{!}{\includegraphics{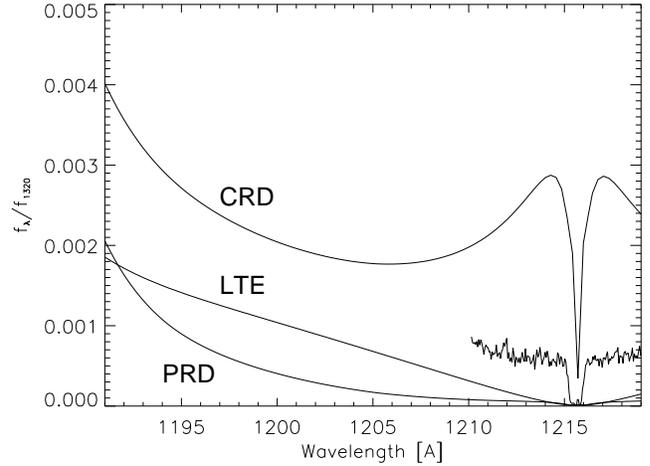}}
%  \rule{0.4pt}{2cm}
  \caption{Predicted L$\alpha$ line profiles normalized to the flux at
    1320\,\AA, calculated with model EXT2 and assuming LTE, complete or
    partial photon redistribution. Vega spectrum similarly normalized is
    also shown in the central part of L$\alpha$.}
  \label{figpcrd}
\end{figure}
%_____________________________________________________

In most applications, one assumes non-coherent scattering in spectral-line
formation and that photons are completely redistributed over the line
profile (CRD approximation). Neither this limit nor the strictly coherent
scattering limit is achieved in stellar atmospheres, requiring to calculate
redistribution functions in some detail. Partial redistribution (PRD)
effects on predicted line profiles turn out to be of importance mainly
for resonance lines (see e.g. Mihalas \cite{book78}) and, particularly,
for L$\alpha$ where scattering is almost coherent (Vernazza et al.
\cite{VAL1}, Hubeny \cite{hubeny80}).

We have computed the L$\alpha$ profile assuming LTE (i.e. no scattering),
CRD or PRD, for model EXT2. Partial redistribution in L$\alpha$ is treated
following Hubeny (\cite{hubeny80}), as implemented in SYNSPEC (Hubeny et al.
\cite{synspec}). Fig.~\ref{figpcrd} shows that the three
line profiles differ markedly. Basically, L$\alpha$ is a scattering
line (i.e. the source function is determined by the mean intensity, $J$,
and the thermal part is very small) and, therefore, the predicted
profile depends critically on the assumed approximation for scattering.
In CRD, photons absorbed in the Doppler core will be reemitted in the 
line wings where they could escape more easily. This contributes to
increase the flux in the wings. The fraction of non-coherent scattering
is small in the PRD approximation resulting in a deeper line profile.
All non-LTE models, but EXT4, predict in CRD a central bump in
L$\alpha$ profile. This central bump is formed at depths where 
H$\alpha$ becomes optically thin. H$\alpha$ photons can escape, resulting
in an increase of the $n=2$ level population, and eventually increasing
the number of L$\alpha$ core photons redistributed to the near-wings.
This bump is therefore a spurious result of the CRD assumption.

None of the three predicted profiles reproduce Vega's spectrum in the
center of L$\alpha$. While we explained the spurious bump predicted in
CRD, the source function in PRD and in LTE is too low to account for the
observed flux. There are two possible ways to improve this situation.
One may either assume a larger fraction of non-coherent scattering
than currently used in the PRD approach, or increase the temperature
at the depth of formation (e.g. model EXT5). The second possibility
will be discussed in Sect.~5, while we leave a detailed study of
partial redistribution to a future paper.

%_____________________________________________________

\subsection{Treating \ion{Fe}{ii} outside LTE}

The final step consists in solving the non-LTE \ion{Fe}{ii} line formation.
At this point, we keep fixed the atmospheric structure and the non-LTE
populations previously calculated to avoid changes in the background
continuum (basically here the L$\alpha$ profile) that may occur with different
model atoms of \ion{Fe}{ii}. In this way, we focus our attention
on the \ion{Fe}{ii} line formation, avoiding indirect effects which could
mask the mechanism responsible of the observed emission lines.

Including \ion{Fe}{ii} in non-LTE calculations has long been a challenge
due to the complexity of this ion, the large number
of energy levels (795 observed in the laboratory, about 11\,000 predicted,
Kurucz \cite{kurlist}) and lines (45\,815 and about 1.26 million,
respectively) involved. Such numbers remain beyond the capacity
of today's computers, and this problem therefore requires some statistical
approach. The idea consists in grouping individual levels with similar
physical properties into superlevels. Individual levels in a given superlevel
share a common non-LTE departure coefficient. We set up different level
groupings  based on two approaches: {\sl (i)\/} grouping by energy and parity,
{\sl (ii)\/} grouping by configuration.

The first approach allows us to tailor freely the number of superlevels.
We have defined energy limits in such a way that all observed levels
are grouped into 80 superlevels. The transitions between superlevels
can include lines in a wide wavelength interval with a complicated
total cross-section. They  are described by Opacity Distribution Functions, 
see details in Hubeny~\& Lanz (\cite{clali1}).

In the second case, the observed levels are grouped into 293
superlevels, according to terms.
We assign the multiplet $gf_{\rm mult}$-value to each
transition between two superlevels, and represent the transition by a single
line with a Gaussian profile. In addition to the case including all
terms, we have set up two simpler test cases. They are
geared toward explaining specifically observed emission lines of two
multiplets (see Sect. 3.1): the doublet $b~^2$H--$x~^2$I$^{\rm o}$, and
the quartet $a~^4$H--$v~^4$F$^{\rm o}$. We keep the model atom as simple as
possible, including only the 10 lowest levels, the two levels of the multiplet,
and those levels that are directly connected to the studied multiplet.
In total, the two models include  40 and 44 \ion{Fe}{ii} superlevels,
respectively.

%______________________________________________________________

\section{Chromospheric emissions in early-A stars?}
\label{chrom}

Emission lines in cool stars are most often the signature of a chromosphere.
A lot of effort has been devoted lately to define the locus of the onset
of convective envelopes that are responsible of chromospheric activity.
Such works as, for example, the recent survey of Marilli et~al.
(\cite{marilli97}), have reached a common conclusion, setting this
limit at spectral type A7. The presence of surface convection and chromospheric
emissions is thus not expected in early A-type stars.

\begin{figure}
  \resizebox{8.8cm}{!}{\includegraphics{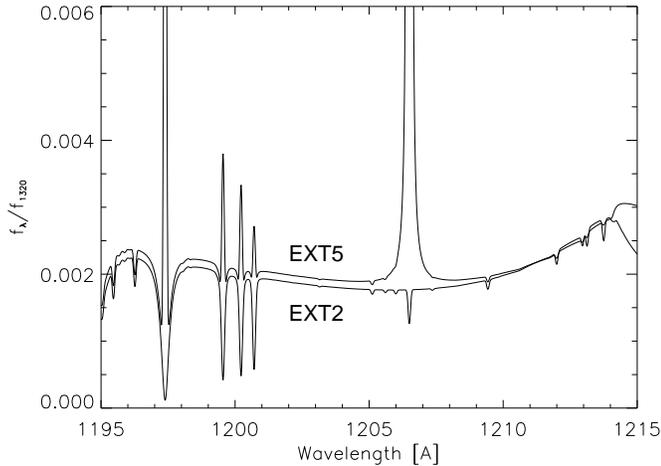}}
%  \rule{0.4pt}{2cm}
  \caption{Theoretical spectra (CRD case) in the blue wing of L$\alpha$
  for models EXT2 and EXT5, without and with an empirical chromospheric
  temperature rise, respectively.}
  \label{figchrom}
\end{figure}

Nevertheless, we decided to investigate the effect of various temperature
stratification on the emergent L$\alpha$ line profile and the \ion{Fe}{ii}
lines, since higher temperatures in the line-forming region might explain
our observations. We have computed a spectrum for each model, EXT1 to EXT5,
in the CRD and PRD cases. Here, we assume LTE for \ion{Fe}{ii} line formation.
Fig.~\ref{figchrom} presents the emergent spectrum of models EXT2 and EXT5
in the CRD case. From these models, four main points emerge:
\begin{itemize}
\item[--] As far as the L$\alpha$ profile is concerned, model EXT2 is
indistiguishable from EXT1. Model EXT5, with a chromospheric temperature
rise starting at $\tau_{\rm R}\simeq 10^{-3.5}$, differs only in the
central part, $\lambda\lambda$\,1214-1217\AA, which emerges from these
hotter layers. Model EXT3 exhibits small differences in the same range. 
\item[--] On the contrary, increasing the temperature in deeper layers
(model EXT4) produces a marked change in the line profile; the calculated
flux becomes about 5 times larger than observed if CRD is assumed; on
the other hand, if PRD is assumed, EXT4 predicts a flux closer
to the observations in the center of L$\alpha$ (compare Fig.~\ref{figpcrd}
and bottom panel of Fig.~\ref{schuster}).
\item[--] \ion{N}{i}, \ion{Si}{ii}, and especially \ion{Si}{iii}
resonance lines turn to strong emissions in models with temperature rises. 
\item[--] Finally, \ion{Fe}{ii} lines are affected only very slightly by
changes in the upper atmosphere. They are sensibly weaker in model EXT4 only.
\end{itemize}

%______________________________________________ 
%______________________________________________ 

Turning the \ion{Fe}{ii} lines into emission requires either high
temperatures in relatively deep layers or/and low continuum source
function (i.e. the PRD approximation). However, other 
lines, like the \ion{Si}{ii} and \ion{Si}{iii} lines, then turn also into
emission. Indeed, \ion{Si}{iii}$\lambda$1206 is observed in emission in
A7 stars (Simon~\& Landsman \cite{simon97}), but not in Vega and Sirius.
Could emission in these resonance lines be masked by interstellar absorption
and mimic the observed narrow lines? This is unlikely, because our models
predict broad emissions, either for the \ion{Si}{iii} line in model EXT5
(see Fig. \ref{figchrom}) or for \ion{Si}{ii} lines in model EXT4
(see bottom panel, Fig. \ref{schuster}), and
narrow interstellar absorption lines cannot mask the emission wings.

From these first models, we therefore conclude 
that the emission lines observed in L$\alpha$ wings do not provide
a compelling evidence for the presence of a chromosphere. The temperature
rise, if any, must occur higher than $\tau_{\rm R}\approx\,10^{-3}-10^{-4}$.
This agrees well with the result of a detailed non-LTE study of 
\ion{Mg}{ii} resonance lines (both in CRD and PRD) by Freire Ferrero et al.
(\cite{freire83}). They found that a limit for a chromosphere in Vega must
be set to  $\tau_{\rm R} < 10^{-4}$ to achieve a good match to the
observations. More definite conclusions based on L$\alpha$ must however await
more detailed PRD-NLTE calculations.

%______________________________________________________________

\section{Non-LTE \ion{Fe}{ii} line formation}

\subsection{Schuster mechanism}

%_____________________________________________________
\begin{figure}
  \resizebox{8.8cm}{!}{\includegraphics{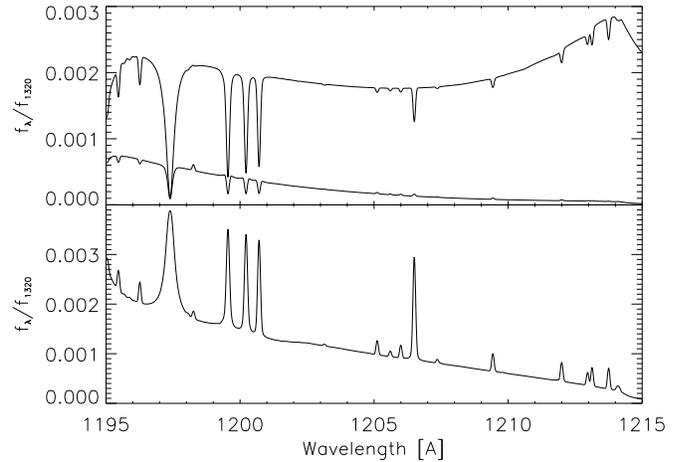}}
%  \rule{0.4pt}{2cm}
  \caption{Emission lines resulting from the Schuster mechanism for two
  models. Top panel: model EXT2 (CRD and PRD approximations); bottom
  panel: model EXT4 (PRD).}
  \label{schuster}
\end{figure}
%_____________________________________________________

Weak lines in L$\alpha$ wings see L$\alpha$ as a strong scattering
local continuum. The interplay between scattering and absorption
may give rise to absorption as well as emission lines, and is know
classically as the Schuster mechanism (see Mihalas \cite{book78}).
Hubeny (\cite{hubeny82}) discussed the case of the different types
of lines in the L$\alpha$ wings of A stars.

Fig. \ref{schuster} illustrates this effect. The top panel shows
that the line profiles are drastically different depending on the
assumption on L$\alpha$ scattering. Here, all lines (except L$\alpha$)
are assumed to be in LTE and, thus, $S^l=B$. The source function
of the local continuum is lower in the PRD case than in the CRD case,
which results in weaker absorption lines in the PRD case. A close inspection
of the figure even shows very weak emissions. Temperature in the
line-forming region is higher in model EXT4, resulting in higher
line source functions and stronger emissions (bottom panel).

The Schuster mechanism may therefore be an important effect to explain
the observed \ion{Fe}{ii} emission lines. Its efficiency depends
however on the temperature gradient in the photosphere as well as on
the details of L$\alpha$ scattering.

\subsection{Radiative interlocking}

Another mechanism that can be responsible for the formation of emission lines
is radiative interlocking. By interlocking, we mean the interaction of two or
more atomic transitions. This is usually important only when one of the
transitions is the dominant process in populating a shared level, and thereby
influences the other transition(s). This can happen for instance 
when the flux in one line is much larger than in an other line.

In the solar spectrum, such situations have been observed. In particular,
an emission line has been observed at 3969.4\,\AA\ 
close to the core of the \ion{Ca}{ii}~H resonance line.
This line has been attributed first to \ion{Eu}{ii}, \ion{Er}{ii}, and
finally to \ion{Fe}{ii}.

Canfield \& Stencel (\cite{canfield76}) suggested that interlocking
usually occurs when the atomic structure is so complex that radiative
interlocking of many weak lines becomes significant, as in rare-earth
elements. On the other hand, metals have a simpler atomic structure,
and emissions are mainly controlled by the presence of strong pumping lines
at shorter wavelengths. They conclude that the \ion{Fe}{ii}$\lambda$3969.4
emission line should be associated with the presence of a chromosphere.

However, non-LTE calculations with a simplified \ion{Fe}{ii}
model atom (Cram et al. \cite{cram80}) indicated that this is not necessarily
the case. \ion{Fe}{ii} emission in the wings of strong resonance lines can
be explained by the same mechanism of radiative interlocking
with a very limited number of transitions.
Their results showed that the line properties depend primarily on the
photosphere characteristics rather than on a chromosphere, just as
for rare-earth elements.

%______________________________________________ 
\begin{figure}
  \hspace{2mm}\psfig{figure=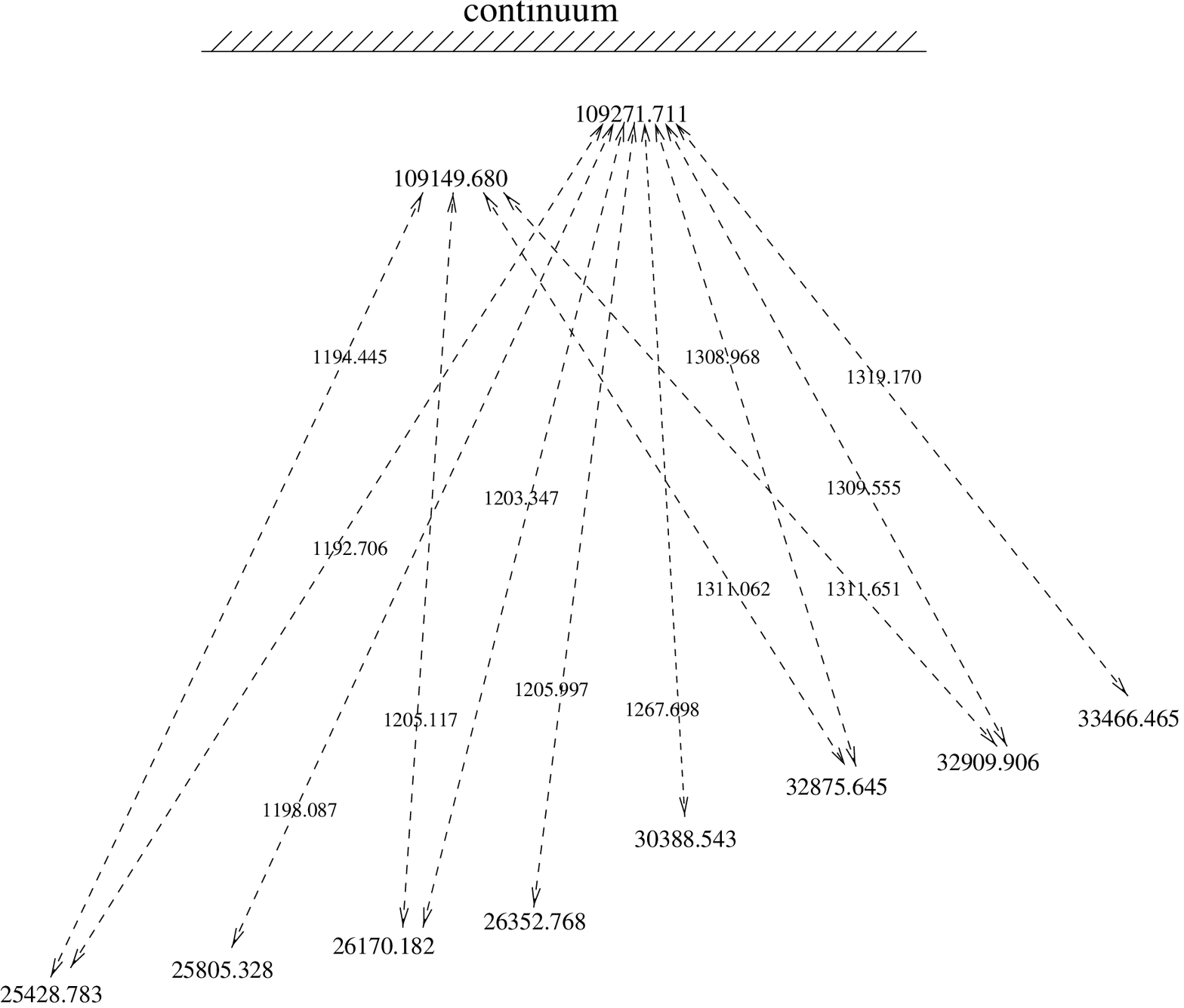,width=8.6cm,angle=-90}
% \resizebox{8.8cm}{!}{\includegraphics{term2.ps}}
%  \rule{0.4pt}{2cm}
  \caption{Most important lines contributing to the population
    of the $x~^2$I$^{\rm o}$ levels, and thus important for the formation
    of the $\lambda$1205.1, 1206.0 lines. The wavelength of each line
    is indicated in \AA, and level energies are in cm$^{-1}$. The ionization
    energy of \ion{Fe}{ii} is 130\,563\,cm$^{-1}$.}
  \label{levcase1}
\end{figure}
%_____________________________________________________

Emission caused by radiative interlocking may be observed in situations
where the flux varies strongly with wavelength, for instance in the wings of
strong resonance lines. The depth of formation between
weak lines in the continuum and weak lines close to the core of resonance
lines can differ by several orders of magnitude in these situations. 
This large flux difference may result in a situation where one transition
is the dominating process in populating the upper energy level.

Since L$\alpha$ presents one of the strongest flux changes between the core
and the continuum, radiative interlocking may therefore be a good
candidate to explain the observed \ion{Fe}{ii} emission lines in Vega
and Sirius. We have tested this idea with the model atmosphere EXT2, and
solving the \ion{Fe}{ii} non-LTE line formation for the different
\ion{Fe}{ii} model atoms described in Sect.~4.4. Here we have still assumed
the CRD approximation to limit the effect of the Schuster mechanism,
and investigate the role of radiative interlocking in \ion{Fe}{ii}.
We will adopt the PRD approximation only in the final comparison (Sect.~6.6)
since we deem it a better approximation to model L$\alpha$.

%_____________________________________________________

\subsection{The ODF approach}

To investigate the non-LTE effects of \ion{Fe}{ii} lines,
we have first used the grouping by energy and parity. In this way, we can
still easily include all individual levels while keeping 
a relative freedom on the number of superlevels. We have adopted a model
with 80 superlevels.

%___________________________________
   \begin{table}
      \caption[]{List of most important levels incorporated in Case~A 
                 (see Fig.~\ref{levcase1}).}
      \label{TabCaseA}
      \begin{tabular}{r l r}
      \hline
       && \\ [-3mm]
       E [cm$^{-1}$]&Term& J \\
      \hline
       && \\ [-3mm]
       25428.783 & $a~^4$G & $ 11/2 $ \\
       25805.328 & $a~^4$G & $ 9/2 $ \\
       26170.182 & $b~^2$H & $ 11/2 $ \\
       26352.768 & $b~^2$H & $ 9/2 $ \\
       30388.543 & $b~^2$G & $ 9/2 $ \\
       32875.645 & $a~^2$I & $ 13/2 $ \\
       32909.906 & $a~^2$I & $ 11/2 $ \\
       33466.465 & $c~^2$G & $ 9/2 $ \\ [2mm]
      109149.680 & $x~^2$I$^{\rm o}$ & $ 13/2 $ \\
      109271.711 & $x~^2$I$^{\rm o}$ & $ 11/2 $ \\
      \hline
      \end{tabular}
   \end{table}
%_____________________________________________________

Some very weak \ion{Fe}{ii} emission lines appeared
in the wings of L$\alpha$. They are predicted however 
much weaker than the observed lines. Moreover, the strongest emission lines
do not coincide with the observed emissions. Many absorption lines are
also predicted weaker than in LTE. This is partly due to overionization,
but may also indicate an increase in the line source function.
This effect was just not strong enough to reproduce the observed emission
lines. A probable reason for this failure is that many levels are grouped
together in the highly-excited superlevels, resulting in some ``dissolution''
of the interlocking effect.

We have tried to improve this situation by splitting some superlevels and
including explicitly a few individual levels, like the $x~^2$I$^{\rm o}$
doublet, which give emission lines in the spectra of Vega and Sirius.
Although this does not result in the expected emissions,
the transitions associated with this doublet were further weakened, suggesting
that radiative interlocking was a contributing factor.

If this approach is most practical to include iron opacities in model
atmospheres, this does not appear to be the best way of grouping levels
for the problem in our hands. 
In a further step, we decided therefore to use a more careful way of grouping
levels by grouping them by configurations. We explore now two test cases where
we focus on two multiplets identified in our spectra. In a final step, we will
again include all \ion{Fe}{ii} energy levels.

%______________________________________________ 
\begin{figure}
  \resizebox{8.8cm}{!}{\includegraphics{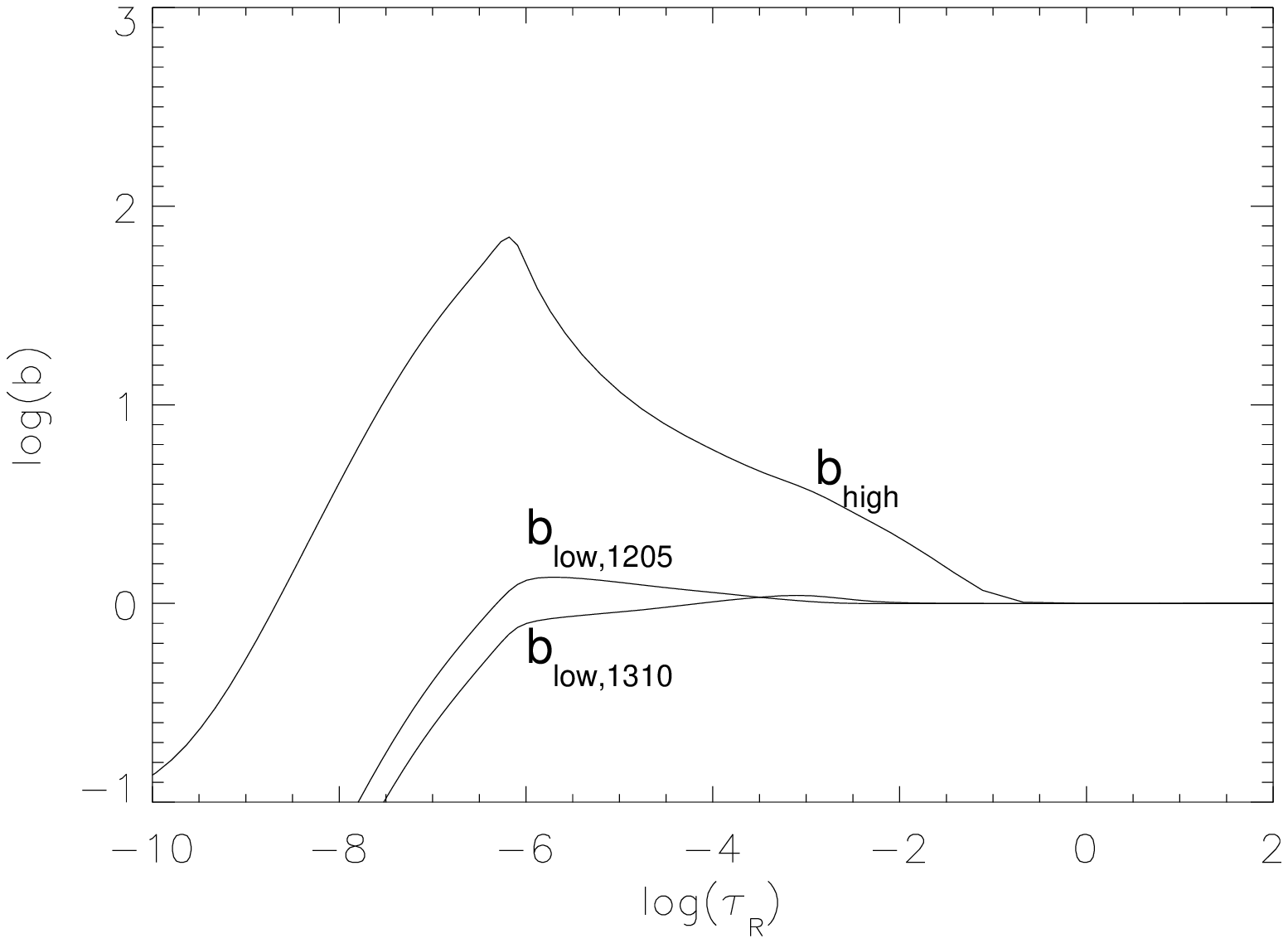}}
  \resizebox{8.8cm}{!}{\includegraphics{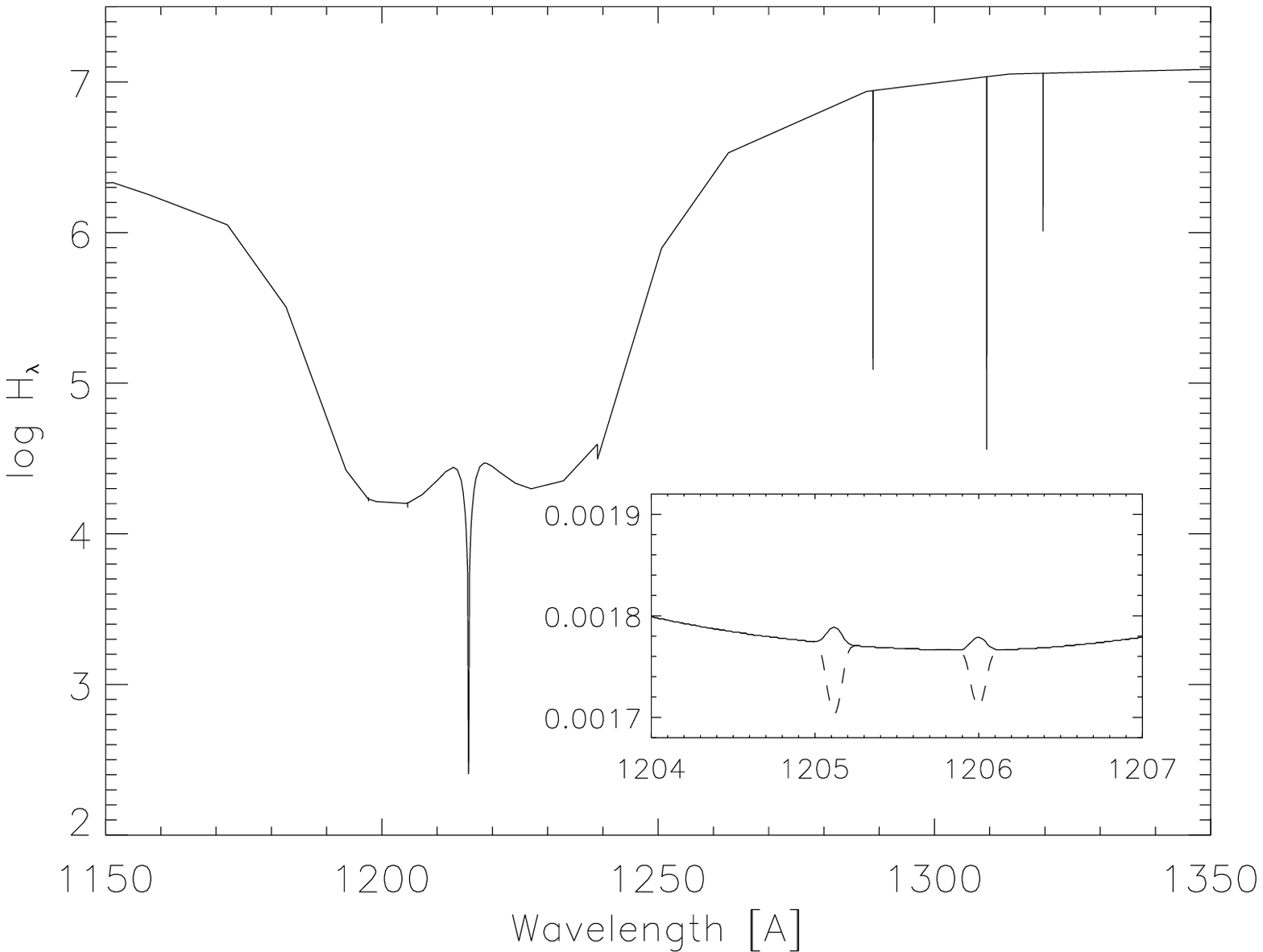}}
%  \rule{0.4pt}{2cm}
  \caption{Results of Case A. Top: non-LTE departure coefficients
     of the $x~^2$I$^{\rm o}$ term and of the two lower levels connected
     to $x~^2$I$^{\rm o}$. Bottom: Emergent spectrum; the \ion{Fe}{ii}
     lines at 1196 and 1205\,\AA. The inset shows the predicted 
     \ion{Fe}{ii} lines compared to the LTE spectrum (dashed line).
     The flux is here normalized at 1320\,\AA.}
  \label{case1a}
\end{figure}
%_____________________________________________________

%______________________________________________ 
\begin{figure}
  \resizebox{8.8cm}{!}{\includegraphics{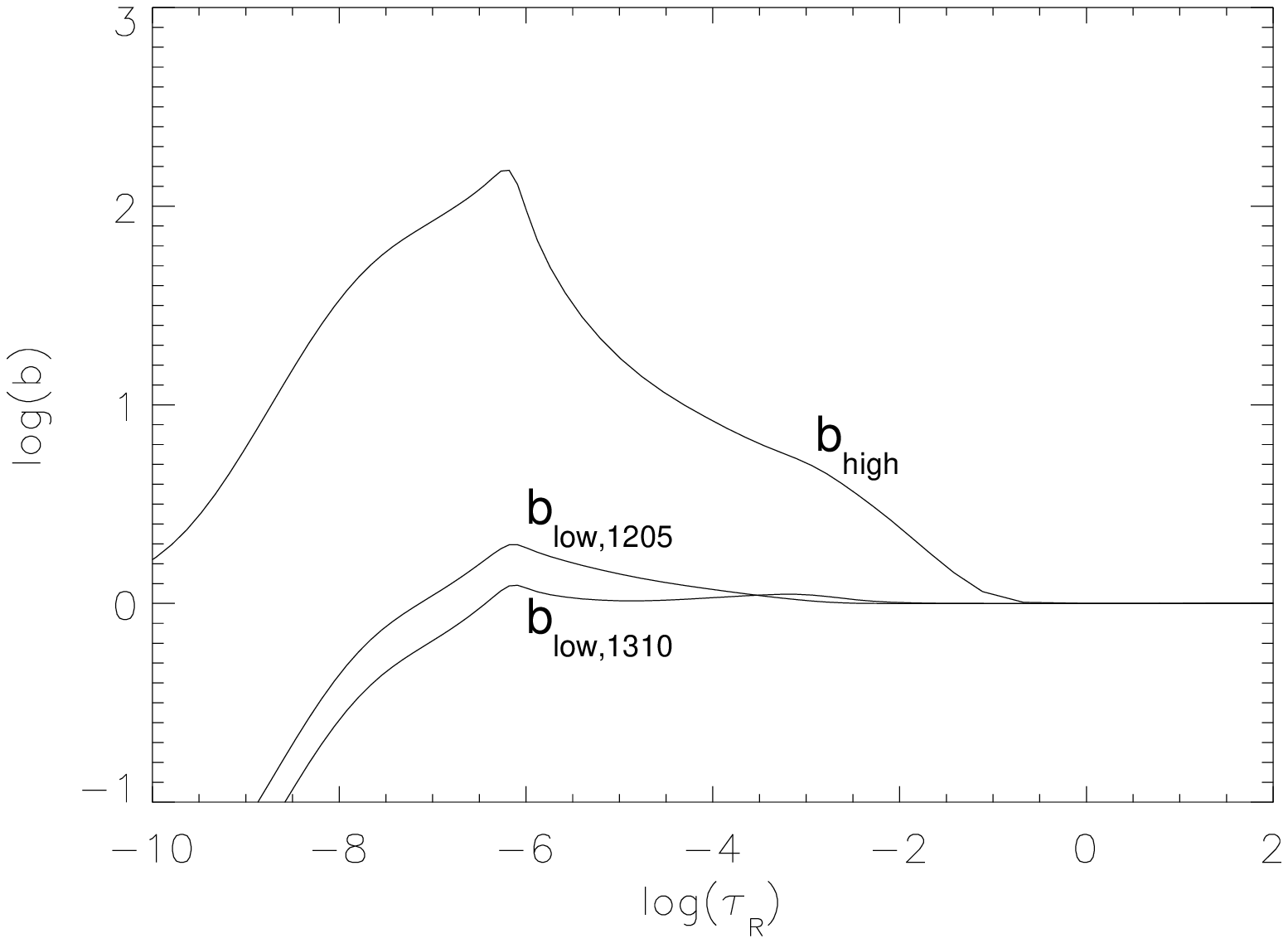}}
  \resizebox{8.8cm}{!}{\includegraphics{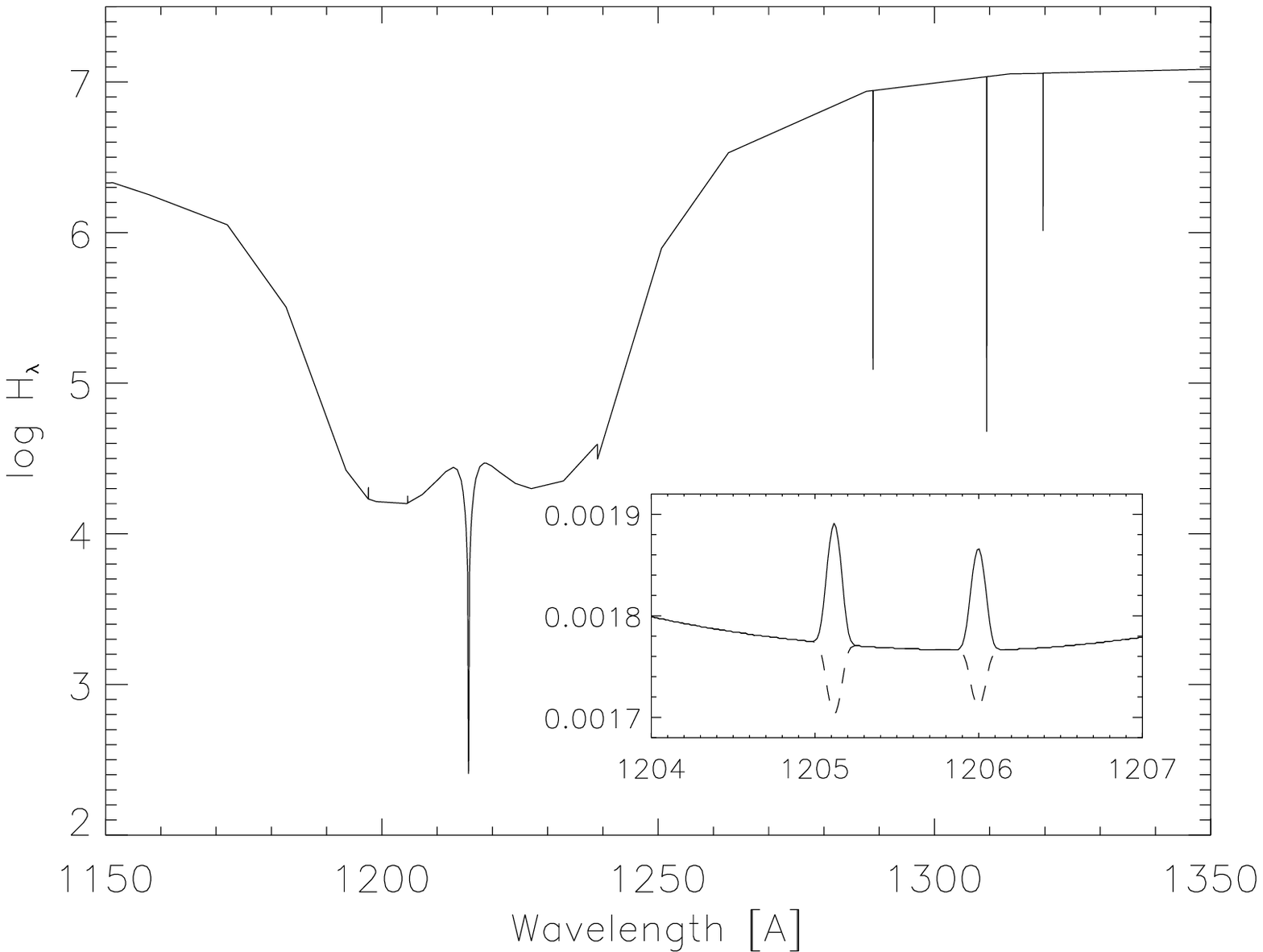}}
%  \rule{0.4pt}{2cm}
  \caption{Same as Fig. \ref{case1a}; new photoionization cross-section
    have been adopted for the $x~^2$I$^{\rm o}$ term. The two
    \ion{Fe}{ii} multiplet lines have turned into stronger emission.}
  \label{case1b}
\end{figure}
%_____________________________________________________

\subsection{Case A: $b~^2$H -- $x~^2$I$^{\rm o}$}

As a first case with \ion{Fe}{ii} grouped by configurations, we have selected
the doublet $b~^2$H -- $x~^2$I$^{\rm o}$
The two lines, at 1205.1 and 1206.0\,\AA, are seen in emission in Vega
and in Sirius. They are thus probably not too sensitive to the detailed
conditions in the stellar atmospheres. The upper levels of these
transitions, 109\,149.7, 109\,271.7$ \, \mbox{cm}^{-1}$,
are also populated by several strong transitions around 1310\,\AA.
Fig.~\ref{levcase1} displays the most important lines connecting to 
these upper levels. Since the flux around 1205$\,$\AA\ is much lower than
around 1310$\,$\AA, we are in a quite favorable situation for radiative
interlocking. The complete system involving the lower and upper levels
of the $\lambda$1205 lines is relatively simple, and we can limit the model
atom to 40 superlevels. The photoionization cross-sections of all 
\ion{Fe}{ii} levels are approximated by an hydrogenic expression.

Our model, assuming atmospheric structure EXT2, shows that indeed
interlocking is important in the formation of these lines.
This can clearly be seen in Fig.~\ref{case1a}, where the non-LTE
departure coefficients of the $x~^2$I$^{\rm o}$, $b~^2$H and $a~^2$I
terms are plotted as a function of the Rosseland
optical depth. The departure coefficient of the upper level has
a value significantly larger than unity in the region of line formation
($\tau_{\rm R}\approx 10^{-3}$). The populations of the two lower levels
show almost no departures from LTE at this depth. The \ion{Fe}{ii}
$\lambda$1205 lines have almost
vanished (Fig.~\ref{case1a}). This stands out in sharp contrast to
the LTE prediction where they are in absorption.
Although the lines are affected as expected, 
the effect falls just short of the observed emission. 

%______________________________________________ 
\begin{figure*}
  \resizebox{12cm}{!}{\includegraphics{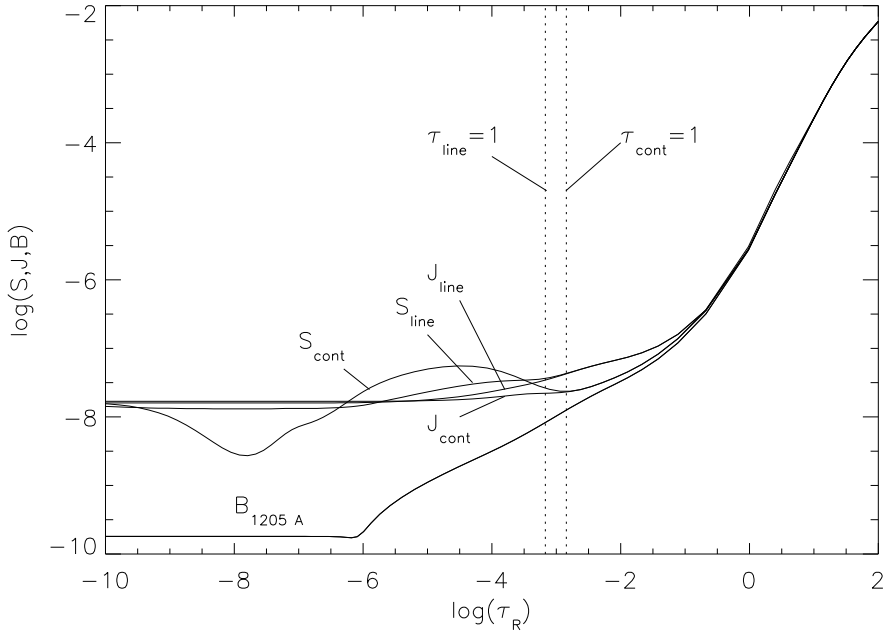}}
  \hfill
  \parbox[b]{55mm}{
    \caption{Source function S, mean intensity J in the line and in the
             local pseudo-continuum, and Planck function B as a function of
             depth. The depth of formation is marked by a departure of S from
             J. The two vertical dotted lines indicate the optical depth
             interval in which the \ion{Fe}{ii}$\lambda$1205 line is formed.}
    \label{jsb}}
\end{figure*}
%_____________________________________________________

The $x~^2$I$^{\rm o}$ energy level is very close to the
continuum. Therefore, ionization and recombination are also important
processes for the population of these levels. Our assumption of an
hydrogenic photoionization cross-section may be unrealistic. S.~Nahar
(priv. comm.) kindly provided us with new theoretical photoionization
cross-sections. Indeed, cross-sections for all $^2$I$^{\rm o}$ energy levels
are notably smaller than the hydrogenic approximation, especially in the
energy regime corresponding to the maximum of flux in Vega. This will reduce
photoionization processes, and increase the non-LTE departure coefficients
of levels close to the continuum. We expect then that lines will turn
more easily into emission.

We have therefore proceeded in changing only the cross-section of level
$x~^2$I$^{\rm o}$ in our Case A. This results in turning the lines into
stronger emission lines (Fig.~\ref{case1b}). The departure coefficients
of the levels differ only slightly from the ones previously calculated with
hydrogenic cross-sections,  but the difference is large enough to
produce the required effect.

Fig. \ref{jsb} displays the run of the source function, mean intensity, and
Planck function with depth. In particular, we see that the
\ion{Fe}{ii}$\lambda$1205 line source function is larger than the
source function of the local continuum (L$\alpha$ wing) in the optical
depth range where the lines are formed. When the 
\ion{Fe}{ii} line becomes optically thin, the source function becomes
larger than the mean intensity $J$ and the Planck function $B$, resulting
in an emission line.

This effect remains small, in particular due to a bump in the continuum
source function. This bump is an artifact of the CRD approximation, 
which leads to a redistribution of L$\alpha$ core photons in
the wings. As noted previously, a better treatment with PRD would prevent
this bump in the source function, 
and increase the contrast between the local continuum and the \ion{Fe}{ii}
lines. This would result in stronger emission lines.

These results show that the interlocking mechanism is capable of explaining
the observed emission lines, by using a relatively simple, isolated
system with a limited number of energy levels. However, the results
remain sensitive to changes in the radiation field, using different
photoionization cross-sections for instance.

\subsection{Case B: $a~^4$D -- $v~^4$F$^{\rm o}$}

A second example is the quartet $a~^4$D -- $v~^4$F$^{\rm o}$, around
1195\,\AA. These lines are observed in emission in Vega, but not in Sirius
(Sect.~3.1). We have incorporated in Case B all levels connected to the
two multiplet levels and the 10 lowest configurations. The total
number of \ion{Fe}{ii} superlevels is 44.

The upper level of this multiplet is not as close
to the ionization limit as Case~A. There are higher levels connected to 
$v~^4$F$^{\rm o}$.
This may result in decreasing the importance of photoionization and
recombination processes. We assume here again a hydrogenic approximation
for the photoionization cross-sections.

In Case B, the resulting emergent spectrum  is not much changed in
comparison to the LTE prediction. Radiative interlocking does not appear to
play an important r\^ole. Although a
larger model atomic system might produce the emission, it seems more likely
to attribute the origin of the emission to the Schuster mechanism in this
case. This would explain why this multiplet is observed in
emission in Vega spectrum only, indicating that the formation of the second
multiplet is more sensitive to the local conditions (radiation field,
temperature, ...) than the Case~A multiplet.

\subsection{Case C: All observed terms}

We must finally examine whether radiative interlocking still produces
emission lines when we incorporate all \ion{Fe}{ii} terms. Many levels
close in energy and numerous collisional transitions might result in
lessening severely the effect of interlocking. Our last
model includes 293 \ion{Fe}{ii} superlevels, 7\,453 radiative transitions,
and about 35\,000 collisional transitions. All photoionization cross-sections
are assumed hydrogenic. The PRD approximation is used to model L$\alpha$.

\begin{figure}
  \resizebox{8.8cm}{!}{\includegraphics{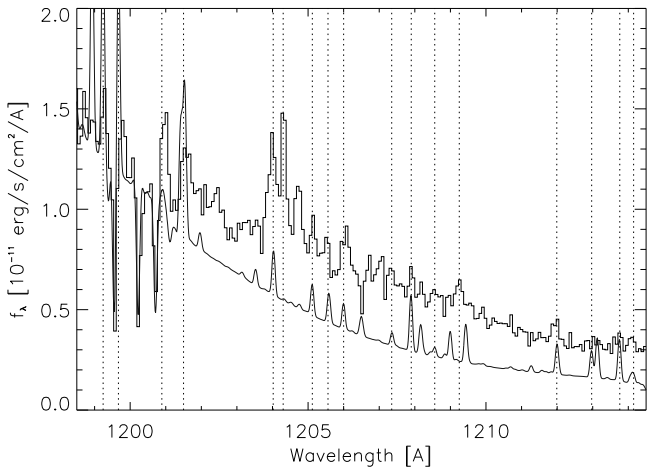}}
%  \rule{0.4pt}{2cm}
  \caption{Emergent spectrum calculated with all \ion{Fe}{ii} levels
  included and grouped into 293 superlevels, compared to Vega GHRS spectrum
  (histogram line).
  Dotted lines show a good correspondence between predicted and observed
  emission lines.}
  \label{vegall}
\end{figure}

Fig.~\ref{vegall} shows a good correspondence
between the predicted and the observed emission lines. The strength
of the emission lines compares also well to the observations.
This demonstrates that the mechanism at the origin
of the observed \ion{Fe}{ii} emission lines is a non-LTE effect, combining
interlocking and the Schuster mechanism. 
Assuming CRD, we also predict
emission lines, but the correspondence is not as good. We believe that
a better correspondence may still be achieved with
realistic photoionization cross-sections for most \ion{Fe}{ii} terms.

%______________________________________________ 
%_____________________________________________________

%______________________________________________________________

\section{Conclusions}

We have reported emission features in the L$\alpha$ profile of Vega
and Sirius-A. These emission lines have been attributed to
\ion{Fe}{ii} and \ion{Cr}{ii} transitions.
The identification appears quite secure
because all the lines of several multiplets appear in emission.

We have built non-LTE model atmospheres with different
assumed temperature structures
in the outer layers and incorporating \ion{Fe}{ii} with different degrees
of sophistication. We found that the emission features cannot be explained
by a chromospheric temperature rise. To produce the observed \ion{Fe}{ii}
emissions, the
temperature would have to increase in relatively deep layers, turning
other lines into emission (e.g. \ion{Si}{ii}, \ion{Si}{iii} lines).
However, we cannot exclude a chromospheric rise in shallower layers
($\tau_{\rm R} \le 10^{-4}$) based on our present observations, in agreement
with earlier results (Freire-Ferrero et~al. \cite{freire83}).

Non-LTE \ion{Fe}{ii} line formation calculations with different model
atoms have demonstrated that some \ion{Fe}{ii} lines can turn into emission
in the wavelength range between 1190 and 1240\,\AA. We stress that
emission lines are predicted {\em only} in this very low flux, central region
of L$\alpha$. This results from the combined effect of the Schuster
mechanism and radiative interlocking. Some highly-excited
levels are overpopulated by transitions occurring in a high-flux region, and
preferentially de-excite in this region near L$\alpha$. This mechanism
explains the similarity of Vega and Sirius spectra.

Differences between the two stars can also be understood with this mechanism.
The higher heavy-element content in Sirius' photosphere results in depressing
the flux, in particular near the flux maximum. The efficiency of the
pumping is thus reduced, yielding generally weaker emissions in Sirius
than in Vega. The details depend on the exact wavelength of the pumping
transitions. The flatter L$\alpha$ profile in Vega is also a consequence
of the different metallicity. Lyman continuum
heating must be more efficient in Vega's case (less heavy-element line
opacity) yielding a somewhat higher temperature in the outer layers and
a higher flux in the central region of L$\alpha$.

We believe that the origin of the \ion{Cr}{ii} emission lines may be
explained by similar mechanisms. While we cannot rule out that
radiative interlocking is also effective in \ion{Cr}{ii},
the difference between Vega and Sirius
points to the Schuster mechanism being the major cause of the emission
in this case. The lines are thus stronger in Sirius due to the
larger chromium abundance, and are simply too weak in Vega to stand
out of the noise. 
 
Although our model atmosphere calculations provide an explanation to
an unexpected observation of emission lines in the spectrum of early
A-type stars, we did not achieve a good fit to the L$\alpha$ profile at this
stage. It seems however likely that the flux observed in the central
region of L$\alpha$ may be explained by increasing somewhat the fraction
of non-coherent scattering in the PRD approximation that we have used.
Matching these observations would require (at least)
non-LTE line-blanketed model atmospheres, treatment of L$\alpha$ in partial
redistribution tuning the ratio between coherent and non-coherent scattering,
and improved, non-hydrogenic \ion{Fe}{ii} photoionization cross-sections.
Such an approach is necessary to gain a deeper insight into the outer layers
of Vega and Sirius, and this certainly deserves further study.

Finally, we did not find emission lines very close to the L$\alpha$ core,
especially in the 0.5\,\AA\  blueward of the central wavelength. This
implies fortunately that we have so far no reason to question the previous
results on the local interstellar cloud
(Bertin et~al. \cite{sirius2}), and on the wind absorption
feature (Bertin et~al. \cite{sirius3}).

\begin{acknowledgements}
We would like to thank the STScI staff for rescheduling Vega and Sirius
observations after an original failure to acquire these two very bright
stars. Especially, we are grateful to Alice Berman who noticed that the plan
for the repeated observations would probably also fail, Dave Soderblom who
proposed another approach that was successful, and Al Schultz who implemented
it. We would like to thank also Ivan Hubeny and C.~S. Jeffery
for many comments and suggestions
which helped to improve this paper. The non-LTE analysis was the topic
of MvN master's thesis in astrophysics at Utrecht University. TL
acknowledges the financial support of the Netherlands Foundation for
Research in Astronomy.
\end{acknowledgements}

\end{document}